# Formation of Nanofoam carbon and re-emergence of Superconductivity in compressed $CaC_6$

Yan-Ling Li,[1,2,*] Wei Luo,[2] Xiao-Jia Chen,[3,4] Zhi Zeng[3], Hai-Qing Lin[5] & Rajeev Ahuja[2,6,*]

[1] *School of Physics and Electronic Engineering, Jiangsu Normal University, 221116, Xuzhou, People's Republic of China*

[2] *Condensed Matter Theory Group, Department of Physics and Astronomy, Uppsala University, P.O. Box 516, SE-751 20 Uppsala, Sweden*

[3] *Center for Energy Matter in Extreme Environments and Key Laboratory of Materials Physics, Institute of Solid State Physics, Chinese Academy of Sciences, Hefei 230031, People's Republic of China*

[4] *Center for High pressure Science and Technology Advanced Research, Shanghai 201203, People's Republic of China*

[5] *Beijing Computational Science Research Center, Beijing 100089, People's Republic of China*

[6] *Applied Material Physics, Department of Materials Science and Engineering, Royal Institute of Technology (KTH), SE-100 44, Stockholm, Sweden*

16 text pages, 4 Figures, 1 Tables

147 words in the abstract, 3097 words in main text.

**Pressure can tune material's electronic properties and control its quantum state, making some systems present disconnected superconducting region as observed in iron chalcogenides and heavy fermion $CeCu_2Si_2$. For $CaC_6$ superconductor ($T_c$ of 11.5 K), applying pressure first $T_c$ increases and then suppresses and the superconductivity of this compound is eventually disappeared at about 18 GPa. Here, we report a theoretical finding of the re-emergence of superconductivity in heavily compressed $CaC_6$. The predicted phase III (space group *Pmmn*) with formation of carbon nanofoam is found to be stable at wide pressure range with a $T_c$ up to 14.7 K at 78 GPa. Diamond-like carbon structure is adhered to the phase IV (*Cmcm*) for compressed $CaC_6$ after 126 GPa, which has bad metallic behavior, indicating again departure from superconductivity. Re-emerged superconductivity in compressed $CaC_6$ paves a new way to design new-type superconductor by inserting metal into nanoporous host lattice.**

Introduction

The metal doped organic compounds, graphite, and solids with π-electrons networks, covering intercalation of alkali metals into buckminsterfullerene solids, metal intercalated graphite intercalation compounds (GICs), and alkali-metal-doped hydrocarbon materials, have recently received much attention in unveiling the new physical properties[1], in particular superconductivity. So far, the highest $T_c$ achieved in this class of materials is 38 K in pressurized $Cs_3C_{60}$[2,3]. In the case of GICs, superconductivity was first observed in $KC_8$ with a transition temperature of 0.15 K[4].

When using calcium atoms to replace the potassium atoms, which brought up to $T_c$ of 11.5 K[5,6], making $CaC_6$ considered to be a template for exploring physical properties and superconducting mechanism of GICs-based superconductors as well as improving $T_c$ value. Pressure, as a powerful tool, is often used to elevate superconductors' $T_c$. Gauzzi *et al* concluded that a large linear increase of $T_c$, from the ambient pressure value 11.5 K up to 15.1 K at 7.5 GPa, followed by a sudden drop to 5K at about 8 GPa due to a first-order structural phase transition[7]. Theoretically, pressured-induced structural phase transition is ascribed to softening phonon modes related to in-plane vibrations from Ca atom[9,10]. It has been found that *R-3m* phase with hexagonal ring transforms into layer orthorhombic *Cmmm* structure with five- and eight-membered rings at 18 GPa[11]. Further, the pressure dependence of $T_c$ studied in ac susceptibility measurements indicated that following an initial increase at the rate +0.39(1) K/GPa, $T_c$ drops abruptly from 15 K to 4K at 10 GPa and no superconducting transition is observed above 2K between 18 and 32 GPa[8]. The repressed superconductivity at about 32 GPa should be correlated to further structural modification, signaling the vicinity of new phase. So far, however, available structural information is scare, which partly hinders theoretical or experimental studies on highly compressed $CaC_6$, since it is indispensable to discern stable structures so as to explore hidden physical property of $CaC_6$ at higher pressure.

## Results

In this study, *ab initio* evolutionary algorithm (EA, USPEX code) [12-14], which has been used to successfully predict high pressure phases of well-known materials, from element solids[15,16] to binary[17,18] or ternary[19] compounds, is chosen to address low enthalpy phases of cold compressed $CaC_6$. Under compression, a novel high pressure phase with nanoporous characteristic, controlling a wide pressure range, presents superconducting feature with considerable $T_c$ of 14.7 K at 78 GPa. The re-emergence of superconductivity has been observed in various systems. Applying magnetic field, Lévy *et al* [20] observed superconductivity in the ferromagnet URhGe. Yuan *et al*[21] observed two disconnected superconducting phases in the pressure-temperature phase diagram of partially germanium-substituted $CeCu_2Si_2$. Very recently, Sun *et al*[22] reported that an unexpected superconducting phase re-emerges in the superconducting iron chalcogenides above 11.5 GPa. Here, re-entrance of superconductivity in cold compressed $CaC_6$ provides new example to discern peculiar physical behavior of material at extreme conditions.

Pressure induces phase transitions from hexagonal to orthorhombic structures. Three orthorhombic high pressure new phases were found. The enthalpies vs. pressure curves of stable structures are presented in Fig. 1 a). We checked stability of $CaC_6$ compared to its element solids using calcium's stable structures reported experimentally and theoretically[23]. No decomposition is observed in our calculations. The *R-3m* structure (referred as phase I) is the most stable structure below 19 GPa, followed by an orthorhombic *C222* structure (19-39 GPa). The *Pmmn* phase (referred as phase II), a good metal (see supplementary Fig. S1), predominates in the pressure range of 39 to 126 GPa. Diamond-like carbon structure is adhered to the phase IV (*Cmcm*) for compressed $CaC_6$ above 126 GPa, showing a bad metallic feature (see supplementary Fig. S2) due to existence of most of saturated $sp^3$ hybridized carbons. Phase II with buckled five- or eight-membered ring's plane has very close enthalpy as of *Cmmm* structure with flat five- or eight-membered ring's plane reported by Csányi *et al*[11]. For these two structures, we cannot say which structure is more stable than other in term of total energy due to their very small structural difference. As pressure increases, buckling characteristics in phase II gradually disappears, making it transform into *Cmmm* symmetry. Phonon spectrum calculations show that both *C222* and *Cmmm* structures are dynamically stable. It is well known that non-hexagonal rings in a perfect graphene can be generated by introducing topological defects[24,25]. Here, external pressure and chemical pre-compression from calcium atoms provide the activation energy to drive di-vacancies defects formed, which leads to graphene layer in phase I transformed into octagonal and pentagonal carbon rings as observed in phase II.

In phase III, there are four types of in-equivalent carbon atoms in the unit cell (referred as C1, C2, C3, and C4, see supplementary Table S1), constructing carbon nanoporous structure. Along y-axis direction, carbon atoms form near quadrangular microporous, in which the size of porous lies in below 2 nm (about 0.61 nm and 0.59 nm at 0 GPa and 39 GPa, respectively). C1 and C2 (four carbon atoms in unit cell) show $sp^3$ hybridization, other two types of carbons C3 and C4 form zigzag chains in the middle of border with two different bonding lengths (1.426 and 1.467 Å at 39 GPa, respectively, see supplementary Table S2), showing mostly $sp^2$ hybridization. Ca atoms stay at center of porous (coordination number 8, with C3 and C4 atoms), forming 1D atomic chain along y-axis direction. At phase transition pressure (39 GPa), the nearest neighbor distance between C and Ca atoms is about 2.31 Å, which is slightly smaller than the value 2.489 Å of Ca-Ca bonding length (i.e, lattice parameter *b*), so there is stronger interaction between $sp^2$ hybridized carbon and calcium because of charge accumulation in the bond region (see supplementary Fig. S3). Strong covalent bonding between carbon atoms ($sp^3$ and $sp^2$ hybridization) bring about its high rigidity, thus having high hardness. To explore its rigidity, we calculate elastic constants (see Table 1) and phonon spectrum of *Pmmn* structure at zero pressure. No imaginary frequency (see supplementary Fig. S4) was observed together with elastic stability in conjuncture with elastic and dynamical stability of carbon host lattice (space group *Cmcm*) at zero pressure (see Table 1 and see supplementary Fig. S5), signaling its rigidity. We also observed a meta-stable high pressure phase ($P2_1/m$) for compressed $CaC_6$, in which carbon nanoporous structure includes armchair-type $sp^2$ carbon chains in the middle of border in porous. For phase IV, under external pressure and chemical pre-compression from Ca sub-lattice, carbon atoms integrate closely into diamond-type property's strip (a total of five layer carbon atoms), while Ca atoms is also rearranged as the bulked plane compared to centered sites of porous in phase III. In diamond-like strips, internal carbon atoms form distorted tetrahedrons with obvious $sp^3$ hybridization (see supplementary Fig. S3), while surface carbon atoms mostly form $sp^2$ hybridization in armchair arrangement way along z-axis direction. The alternate carbon 3D framework and metal layers have an AαBβ arrangement, where the A and B represent the metal layers and α and β the diamond-like strip layers. The formed covalent three dimensional quasi-layer network (Ca layer-diamond-like layer-Ca layer- diamond-like layer) in phase IV largely enhances its incompressibility and hardness in comparison with phase III. Using suggested formula by Šimůnek and Vackář[26], the estimated hardness is 34 GPa for *Pmmn* CaC$_6$ at zero pressure, 45 GPa for *Pmmn* CaC$_6$ at 39 GPa, and 91 GPa for *Cmcm* CaC$_6$ at 126 GPa.

The atomic arrangements of competing structures are shown in Fig. 1 c). The nearest neighbor distance between carbon atoms is 1.444 Å for the *R-3m* phase at 0 GPa, 1.417 Å for the *C222* phase at 19 GPa, 1.426 Å for the *Pmmn* phase at 39 GPa, and 1.460 Å for *Cmcm* phase at 126 GPa, which are close to the atomic distances in graphite. Interestingly, we found that only small change is observed for these distances as pressure increase, indicating very stiff chemical bonding between carbon atoms. For phase III from zero pressure to 39 GPa, carbon-carbon length only decreases 3~4%. The stiffness of C-C bonds corresponds to high vibration frequencies in phonon dispersions (discussed later).

Carbon nanoporous structure is very important because of its potential applications in many fields, such as H$_2$ absorption, nanodevices, supercapacitors and lithium ion batteries, as well as chemical and bio-sensors. Structurally, the carbon foam, a three-dimensional porous including both *sp*$^2$ and *sp*$^3$ hybridized atoms, can be thought of as assembled from graphene planes interconnected rigidly with one another at a specific angle, forming a linear chain of *sp*$^3$ bonded atoms along the junction[27]. Previously, supposed hybrid carbon foam structures with a mixed *sp*$^2$/*sp*$^3$ bonding character[28,29] have received much less attention for lack of direct experimental observation. Recently, Tománek suggested carbon foam may form under non-equilibrium conditions near gain boundaries of a carbon-saturated metal substrate[30]. Here, (2, 2) zigzag carbon foam observed in phase III of $CaC_6$ points out new easy-realized way experimentally via powerful high pressure tool. Electronic structure calculation shows that (2, 2) zigzag carbon foam is a semiconductor with narrow band gap (~0.49 eV, see supplementary Fig. S6). In view of nanoporous scientific and technological importance, we in detail explore host lattice's structural behavior (carbon lattice in phase *Pmmn*) at zero pressure. Removing calcium from lattice, relaxed host carbon sublattice shows higher symmetry (parent group *Cmcm*), which attributes to C1 and C2 as well as C3 and C4's symmetric degeneracy. For Ca guest sublattice, it forms body-centered tetragonal lattice with *Immm* symmetry. When adding calcium atoms into host lattice, interaction between Ca and C leads to lattice's symmetry decrease. Most exciting is that carbon nanoporous structure relaxed remains stable mechanically and dynamically at zero pressure (see Table 1 and supplementary Fig. S5). At zero pressure, carbon foam equilibrium structural parameters are listed in supplementary Table S1. The densities, carbon-carbon bond lengths, and carbon-carbon bond angles of carbon foam and phase III of $CaC_6$ as well as experimental values of graphite and diamond are listed in supplementary Table S2. The foam contains two-thirds 3-fold coordinated carbon atoms, labeled *$sp^2$*, and one-third 4-fold coordinated carbon atoms, labeled *$sp^3$*. The mass density of the optimized foam structure, $\rho$=2.589 g/cm$^3$, lies in-between the experimental values for graphite and diamond (see supplementary Table S2), which is highest density among reported carbon foams so far. Our carbon foam has peculiar structural behavior. There are three types of covalent bonds, $sp^2$-$sp^2$, $sp^2$-$sp^3$, and $sp^3$-$sp^3$ and five types of bond angles, $sp^2$-$sp^2$-$sp^2$, $sp^2$-$sp^3$-$sp^2$, $sp^2$-$sp^3$-$sp^3$, $sp^3$-$sp^2$-$sp^3$, and $sp^3$-$sp^3$-$sp^3$, as presented in Table supplementary S2, which leads to non-hexagonal cross-section at junction. From supplementary Table S2, it can be seen that $sp^2$-$sp^2$ bond length in carbon foam is shorter than that in graphite, again indicating its stronger rigidity, which differs from previous reported carbon foam[27,29]. The size of nanoporous is often measured by its border length. For (2, 2) zigzag carbon foam observed here, it has 3.925 Å of width of tube-wall. The pore diameters (diagonal lengths) of nanoporous are 5.935 Å and 5.408 Å. Further, nanoporous's size can be modified by change numbers of carbon six-membered rings of borders, which helps to tune specific surface area and band gap so as to facilitate requisite applications in purification and separation, adsorption, and catalysis[30]. This shape modification of nanoporous has been observed in other compressed GICs $KC_8$ and $LiC_6$. Using Li as guest, we obtained (2, 2) nanoporous and triangular nanoporous, and using K, two kinds of nanoporous discovered (see supplementary Fig. S7). We can expect that more nanoporous solids can be exposed by other metal guest (such as Sc, Ba, Cs, and Yb *et al*). Structure decides property, so further studies remains desirable to uncover structural evolution behavior of GICs under compression. Also we found that some carbon nanoporous systems with peculiar shape observed here present negative linear compressibility[30]. To add calcium atoms into (2, 2) zigzag carbon foam brings about porous distortion due to strong hybridization between calcium $d$-electrons and $sp^2$ hybridized carbon p-electrons (see supplementary Fig. S1). As is shown in supplementary Table S2, there are five bond lengths and nine bond angles for carbon atoms. Carbon-carbon bond is slightly increased in length because of calcium atomic insertion but presents strong rigidity.

The small density of states (DOS) value at Fermi level for phase II exactly explains disappearance of $T_c$ of $CaC_6$ at the pressure range from 18 to 32 GPa experimentally[8]. Phase III has high DOS value at Fermi level compared with that of phase I (see supplementary Fig. S8) stimulates our intense interest on its potential superconductivity, since the electronic DOS at the Fermi level sensitively controls $T_c$. Energy bands of phase III at 39 GPa were shown in Fig. 2 (a). There are four energy bands, labeled as 1, 2, 3, and 4, crossing the Fermi level along several high symmetry directions in BZ. The most interesting feature is the appearance of a small region (TY direction in BZ) of one two-fold degenerating flat band lying just at the Fermi level around the Y point. The occurrence of flat and steep slopes near the Fermi level, which is a favorable condition for enhancing Cooper pair formation, has been thought to be essential to superconductivity[31,32]. The calculated DOS shows that a strong hybridization between the Ca-$d$ electrons and $sp^2$ carbon-$p$ electrons (C3 and C4-$p$ electrons) appears in the valence band from -2.78 to 0 eV and in the conduction band from 0 to 4 eV. Electrons from $sp^3$ carbon (C1 and C2)-$p$ electrons construct $sp^3$ chemical bonding, lying mostly valence band from -16eV to -2.78eV. Dominant contributions to the DOS near the Fermi level come from the Ca-$d$ electrons and $sp^2$ carbon-$p$ electrons (C3 and C4), whereas the Ca-$p$, C1-$p$, and C2-$p$ electrons make minor contributions to the DOS at the Fermi level (see supplementary Figure S1). Figure 2 (b) shows the Fermi surface (FS) of the *Pmmn* structure at 39 GPa. The labeled numbers (1-4) correspond to the labeled energy bands shown in Fig. 2 (a). The FS consists of two electron-hole oblate tube with an '8'-like section along the S-Y direction (all four bands), two hole tubes (bands 3 and 4) near the S-X-U-R plane, and two hole pockets (band 2, Γ-Z direction). Obvious near parallel pieces of the Fermi surface exists, which favors to enhance the electron-phonon coupling. Both the Fermi pockets and the obvious Fermi nesting create strong electron-phonon interactions in $CaC_6$. By further analyzing the wave functions of the four bands at the high symmetry points, we conclude that electron-like Fermi surfaces arise mainly from the Ca-$d_z^2$ and Ca-$d_x^2$ states, whereas hole-like Fermi surfaces attributes to the $p_x$ and $p_z$ states from C3 and C4.

**Discussions**

Phonon dispersions and the partial phonon density of states (PPHDOS) of phase III are shown in Fig. 3. The flat bands observed along S-X-U-R direction may serve as evidence of its local two-dimensional feature. Many soften modes are observed along the Γ-Z-T-Y and Γ-R directions (see also supplementary Fig. S9). As seen in the PPHDOS (in the right panel of Fig. 3), the calcium atomic vibrations mostly contribute to the phonon dispersions below 385 cm$^{-1}$. The *sp*$^3$ carbon (C1 and C2) atomic vibrations mainly dominate narrow frequency range around 1200 cm$^{-1}$. The phonon dispersions between 385 and 889 cm$^{-1}$ as well as above 1226 cm$^{-1}$ are mainly due to the contributions of the *sp*$^2$ carbon (C3 and C4) atomic vibrations.

Using the phonon linewidth $\gamma_{\mathbf{Q}\nu}$, we can identify the contribution to the electron-phonon interaction parameter λ from each mode ($\lambda_{\mathbf{Q}\nu}$) based on the relation $\lambda_{\mathbf{Q}\nu} = \frac{\gamma_{\mathbf{Q}\nu}}{\pi N(0) \omega_{\mathbf{Q}\nu}^2}$. Here, $N(0)$ is the density of states at Fermi surface. We notice that large values of $\lambda_{\mathbf{Q}\nu}$ lie at the high symmetry points Γ, S, X, U, and R. For instance, we found that at the R point, $\lambda(\omega)$ originates mainly from contributions of the vibration modes at the frequency ranges of 242 to 305 cm$^{-1}$. The calculated spectral function $\alpha^2 F(\omega)$ and integrated $\lambda(\omega)$ of *Pmmn* CaC$_6$ at 39 GPa are plotted in Fig. 4 a). The vibrations below 385 cm$^{-1}$ provide the major contribution to λ (about 0.45). The low-frequency phonons (below 385 cm$^{-1}$), which mostly involve the Ca and *sp*$^2$ carbon (C3 and C4) atoms yield 54.5 % of the total λ value. The phonons between 385 and 889 cm$^{-1}$ mainly from *sp*$^2$ carbon atoms, contribute to 32.4 % of total λ. These results indicate that the calcium and *sp*$^2$ carbon atoms in the *Pmmn* structure dominate superconductivity in CaC$_6$, due to the prominent contributions to the electron-phonon interaction. Phonons from the Ca and *sp*$^2$ carbon atoms together with the electrons from the Ca-*d* and *sp*$^2$ C-*p* states provide the strong electron-phonon coupling necessary for strong superconductivity in *Pmmn* CaC$_6$.

The Allen and Dynes modified formula[34] was used to estimate the superconducting transition temperature $T_c$ from the value of λ determined above. Taking a typical value of 0.115 for $\mu^*$ along with the calculated $\omega_{\log}$ of 642 cm$^{-1}$, we obtained a $T_c$ of 3.48 K for *Pmmn* CaC$_6$ at 39 GPa. Applying pressure to *Pmmn* CaC$_6$ clearly leads to λ first enhanced remarkably, then weaken and thus $T_c$ corresponding responded (see supplementary Table S3). The calculated highest value of $T_c$ is up to considerate 14.7 K (at 78 GPa). At 117 GPa, $T_c$ is only 4.74 K. The calculated decrease of $T_c$ agrees with occurrence of another new phase (*Cmcm*). The change of $T_c$ with increasing pressure in $CaC_6$ is summarized in Fig. 4 b). Phase III has consistent superconducting mechanism with phase I. Re-emerged superconductivity in $CaC_6$ is closely related to structural transformation. Carbon nanofoam centered calcium atom chains in phase III brings about good metallic feature and strong electron-phonon coupling and thus superconductivity. Pressure-induced atomic hierarchical arrangement in phase IV, presenting diamond-like carbon strips (mostly *sp*$^3$ carbon) and bucked calcium atomic layers, decreases its conductivity and thus represses superconductivity. By tuning electronic order (orbital or spin), pressure or magnetic field-driven the re-emergence of superconductivity has been observed in various systems including ferromagnet URhGe[20], heavy fermion compound $CeCu_2Si_2$[21], and iron chalcogenides[22].

To conclude, for compressed $CaC_6$, carbon atom's arrangement is transformed from preceding two-dimensional (2D) layer into three-dimensional (3D) framework, while calcium ions themselves from 3D network turn into 2D buckled plane. A novel high pressure phase (*Pmmn*, two molecules/cell) with nanofoam carbon shows superconducting behavior due to enhanced electronic distribution at Fermi level as well as strong electron-phonon coupling, capturing critical superconducting temperature of 14.7 K at 78 GPa. The pressure-induced re-emergence of superconductivity in $CaC_6$ signals a rich physics for graphite intercalation compounds under high compression. Although compressed $LiC_6$ holds nanoporous carbon in its stable high pressure phases but it does not show the superconducting behavior. Therefore, further theoretical and experimental efforts are well worth to discern physical nature of GICs under compression.

**Methods**

Stable structures of GICs under cold compression were searched using evolutionary algorithm in combined with VASP code[35] based on density functional theory within the generalised gradient approximation with the exchange-correlation functional of Perdew Burke Ernzerhof[36], employing the projected augmented wave (PAW)[37,38] method where $2s^22p^2$ and $3s^23p^64s^2$ are treated as valence electrons for C and Ca atoms, respectively. For the crystal structure searches, we used a plane-wave basis set cutoff of 700 eV and performed the Brillouin zone integrations using a coarse k-point grid. The most interesting structures were further relaxed at a higher level of accuracy with a basis set cutoff of 1000 eV and a k-point grid of spacing $2\pi \times 0.018$ Å$^{-1}$. Iterative relaxation of atomic positions was stopped when all forces were smaller than 0.001 eV/ Å. The lattice dynamic and superconducting properties of *Pmmn* CaC$_6$ were calculated by the Quantum ESPRESSO package[39] using Vanderbilt-type ultra-soft potentials with cutoff energies of 55 Ry and 500 Ry for the wave functions and the charge density, respectively. The electronic Brillouin zone (BZ) integration in the phonon calculation was based on a $12 \times 16 \times 12$ of Monkhorst-Pack k-point meshes. The dynamic matrix was computed based on a $2 \times 4 \times 2$ mesh of phonon wave vectors. The electron-phonon coupling was convergent with a finer grid of $48 \times 64 \times 48$ **k** points and a Gaussian smearing of 0.01 Ry.

## Acknowledgements

This work was supported by the NSFC (11047013), the Priority Academic Program Development of Jiangsu Higher Education Institutions (PAPD), and Jiangsu Overseas Research & Training Program for University Prominent Young & Middle-aged Teachers and Presidents. We thank Swedish Research Council (VR) for financial support. X.-J.C. was supported as part of EFree, an Energy Frontier Research Center funded by the Department of Energy, Office of Science, Office of Basic Energy Sciences, under Award DESC0001057. Part of the calculations was performed at the Center for Computational Science of Hefei Institutes of Physical Science of Chinese Academy of Sciences, the ScGrid of Supercomputing Center and Computer Network Information Center of the Chinese Academy of Science, and the Swedish National Infrastructure for Computing (SNIC).

## Author contributions

YL and RA designed research. YL performed research. All authors analyzed data. YL, XJC, HQL and RA wrote the paper.

## Additional information

Competing financial interests: The authors declare no competing financial interests.

**Figure legends**

**Figure 1 (a) Relative enthalpy per chemical formula unit of compressed CaC$_6$ as a function of pressure, referenced to the *Pmmn* structure. (b) Structural phase transitional diagram under cold pressure. (c) Atomic arrangement of the R-3m, C222, *Pmmn*, and *Cmcm* structures of CaC$_6$.** The large ball represents the Ca atom. There are one, two, four, and three in equivalent carbon atoms in the *R-3m*, *C222*, *Pmmn*, and *Cmcm* structures, respectively. Under pressure, six-membered ring re-arranges into five- and eight-rings. Further, carbon polymerizes into three dimensional networks. Carbon mixing hybridization of *sp$^3$+sp$^2$* is observed in *Pmmn* and *Cmcm* phases. For *Pmmn* structure, *sp$^2$* hybridized bonds are linked by *sp$^3$* bonds, forming nanoporous structure, which is centered by calcium atoms chains. For *Cmcm* structure, the arrangement of carbon atoms is transformed into *sp$^3$* framework structure (diamond-like behavior) with *sp$^2$* bonds as borders and calcium ions have rearranged into 2D buckled plane.

**Figure 2 Electronic band structure along high symmetry lines of the Brillouin zone (top) and the Fermi surface (bottom) of CaC$_6$ in the *Pmmn* phase at 39 GPa.** The energy bands crossing the Fermi level are labelled as 1, 2, 3, and 4, respectively. The projected Ca-*d*, and C3- and C4-*p* electron characters are denoted by red circle and blue square, respectively.

**Figure 3 Phonon dispersion along the high-symmetry directions of the Brillouin zone (left panel) and the partial phonon density of states (PPHDOS) (right panel) of the *Pmmn* CaC$_6$ at 39 GPa.**

**Figure 4 (a) The Eliashberg phonon spectral function $\alpha^2 F(\omega)$ and electron-phonon integral $\lambda(\omega)$ of the *Pmmn* CaC$_6$ structure at 39 GPa. (b) Superconducting transition temperature versus pressure for CaC$_6$.**

*Supplementary information*

# Formation of Nanofoam carbon and re-emergence of Superconductivity in compressed CaC$_6$

Yan-Ling Li,[1,2,*] Wei Luo,[2] Xiao-Jia Chen,[3,4] Zhi Zeng,[3], Hai-Qing Lin[5] & Rajeev Ahuja[2,6,*]

[1] School of Physics and Electronic Engineering, Jiangsu Normal University, 221116, Xuzhou, People's Republic of China

[2] Condensed Matter Theory Group, Department of Physics and Astronomy, Uppsala University, P.O. Box 516, SE-751 20 Uppsala, Sweden

[3] Center for Energy Matter in Extreme Environments and Key Laboratory of Materials Physics, Institute of Solid State Physics, Chinese Academy of Sciences, Hefei 230031, People's Republic of China

[4] Center for High pressure Science and Technology Advanced Research, Shanghai 201203, People's Republic of China

[5] Beijing Computational Science Research Center, Beijing 100089, People's Republic of China

[6] Applied Material Physics, Department of Materials Science and Engineering, Royal Institute of Technology (KTH), SE-100 44, Stockholm, Sweden

**This file includes Three Tables and nine Figures.**

**Supplementary Table S1. Structures of stable phases of CaC$_6$ and carbon foam.**
Only the fractional coordinates of symmetry inequivalent atoms are given. For *C222* phase, under pressure, it is easy to transform into *Cmmm* symmetry, which is verified by total enthalpy calculation.

| Pressure (GPa) | Space group (No.) | Lattice parameters $(a,b,c,\alpha,\beta,\gamma)$ (Å, °) | | | Atomic fractional coordinates | | | |
|---|---|---|---|---|---|---|---|---|
| 0 | *R-3m* | 5.1683 | 5.1683 | 5.1683 | Ca 4*i* | 0.0000 | 0.0000 | 0.0000 |
|  | (164) | 49.7 | 49.7 | 49.7 | C1 6*g* | 0.1667 | 0.8333 | 0.5000 |
| 19 | *C222* | 9.1472 | 3.6947 | 3.5964 | Ca 2*a* | 0.0000 | 0.0000 | 0.0000 |
|  | (29) | 90 | 90 | 90 | C1 4*f* | 0.4213 | 0.0000 | 0.5000 |
|  |  |  |  |  | C2 8*l* | 0.1751 | 0.8082 | 0.4999 |
| 39 | *Pmmn* | 5.8904 | 2.4893 | 6.4903 | Ca 2*b* | 0.0000 | 0.5000 | 0.7425 |
|  | (59) | 90 | 90 | 90 | C1 2*b* | 0.0000 | 0.5000 | 0.3159 |
|  |  |  |  |  | C2 2*a* | 0.0000 | 0.0000 | 0.1766 |
|  |  |  |  |  | C3 4*f* | 0.1994 | 0.0000 | 0.0286 |
|  |  |  |  |  | C4 4*f* | 0.2791 | 0.0000 | 0.5537 |
| 126 | *Cmcm* | 2.3828 | 16.1113 | 4.0533 | Ca 4*c* | 0.0000 | 0.4874 | 0.2500 |
|  | (63) | 90 | 90 | 90 | C1 8*f* | 0.0000 | 0.1044 | 0.4303 |
|  |  |  |  |  | C2 8*f* | 0.0000 | 0.2705 | 0.5647 |
|  |  |  |  |  | C3 8*f* | 0.0000 | 0.3551 | 0.4371 |
| 0 | C$_6$:*Cmcm* | 5.9354 | 6.3121 | 2.4675 | C1 4*c* | 0.0000 | 0.5716 | 0.2500 |
|  | (63) | 90 | 90 | 90 | C2 8*g* | 0.2949 | 0.2163 | 0.2500 |

**Supplementary Table S2.** The bond lengths and bond angle of carbon nanofoam ($C_6$, space group *Cmcm*), *Pmmn*, *Cmcm* phases of compressed $CaC_6$ compared to experimental data of graphite and diamond.

| Structure | Density(g/cm$^3$) | Bond length (Å) | | Bond angle ($^0$) | |
|---|---|---|---|---|---|
| Graphite | 2.27 | $sp^2$-$sp^2$ | 1.421 | $sp^2$-$sp^2$-$sp^2$ | 120.00 |
| diamond | 3.54 | $sp^3$-$sp^3$ | 1.545 | $sp^3$-$sp^3$-$sp^3$ | 109.47 |
| 0 GPa | | | | | |
| nanofoam | 2.589 | $sp^2$-$sp^2$ | 1.410 | $sp^2$-$sp^3$-$sp^2$ | 106.24 |
| $C_6$ (*Cmcm*) | | $sp^2$-$sp^3$ | 1.522 | $sp^3$-$sp^3$-$sp^3$ | 107.55 |
| | | $sp^3$-$sp^3$ | 1.529 | $sp^3$-$sp^3$-$sp^2$ | 110.77 |
| | | | | $sp^3$-$sp^2$-$sp^2$ | 118.93 |
| | | | | $sp^2$-$sp^2$-$sp^2$ | 120.11 |
| *Pmmn* | 3.499 | $sp^2$-$sp^2$ | 1.470,1.513 | $sp^2$-$sp^3$-$sp^2$ | 101.70,112.70 |
| | | $sp^2$-$sp^3$ | 1.582,1.620 | $sp^3$-$sp^3$-$sp^3$ | 107.43 |
| | | $sp^3$-$sp^3$ | 1.596 | $sp^3$-$sp^3$-$sp^2$ | 109.14,111.94 |
| | | | | $sp^3$-$sp^2$-$sp^2$ | 117.00,118.73 |
| | | | | $sp^2$-$sp^2$-$sp^2$ | 116.46,122.24 |
| 39 GPa | | | | | |
| *Pmmn* | 3.914 | $sp^2$-$sp^2$ | 1.426,1.467 | $sp^2$-$sp^3$-$sp^2$ | 101.59,113.92 |
| | | $sp^2$-$sp^3$ | 1.520,1.552 | $sp^3$-$sp^3$-$sp^3$ | 108.02 |
| | | $sp^3$-$sp^3$ | 1.538 | $sp^3$-$sp^3$-$sp^2$ | 108.69,111.80 |
| | | | | $sp^3$-$sp^2$-$sp^2$ | 117.07,119.01 |
| | | | | $sp^2$-$sp^2$-$sp^2$ | 116.05,121.50 |
| 126 GPa | | | | | |
| *Cmcm* | 4.788 | $sp^2$-$sp^2$ | 1.462 | $sp^3$-$sp^2$-$sp^2$ | 111.58 |
| | | $sp^2$-$sp^3$ | 1.461 | $sp^2$-$sp^3$-$sp^3$ | 111.58 |
| | | $sp^3$-$sp^3$ | 1.458,1.460 | $sp^3$-$sp^3$-$sp^3$ | 107.19, 109.41 |
| | | | 1.502,1.517 | | 110.78, 111.06 |
| | | | | $sp^2$-$sp^3$-$sp^2$ | 109.26 |
| | | | | $sp^3$-$sp^2$-$sp^3$ | 109.26 |

[1] Kittel, C. *Introduction to solid state physics* (8th edition, 2004).

**Supplementary Table S3.** The calculated values of the density of states at the Fermi level $N(0)$, phonon logarithmic average $\omega_{\log}$, electron-phonon interaction $\lambda$, and superconducting transition temperature $T_c$ of $CaC_6$ for the *Pmmn* structure at different pressures. $\mu^*$ is the Coulomb pseudopotential parameter.

| Space group | $P$ (GPa) | $N(0)$ (/Ry) | $\omega_{\log}$ (K) | $\lambda$ | $T_c$ ($\mu^*$=0.11) | $T_c$ ($\mu^*$=0.115) | $T_c$ ($\mu^*$=0.12) |
|---|---|---|---|---|---|---|---|
| *Pmmn* | 39  | 15.645 | 642 | 0.445 | 3.87 | 3.48 | 3.12 |
|        | 51  | 15.58  | 663 | 0.587 | 12.7 | 11.9 | 11.2 |
|        | 78  | 15.378 | 623 | 0.638 | 15.4 | 14.7 | 13.9 |
|        | 117 | 14.830 | 728 | 0.461 | 5.23 | 4.74 | 4.28 |

# Figure captions

**Supplementary Fig. S1. Projected density of states of *Pmmn* structure at phase transition pressure.** One can see that there is strong hybridization between Ca-$d$ electrons and $p$ electrons of C3 and C4 (forming $sp^2$ hybridized zigzag chains along y-axis direction). $sp^3$ hybridized C1 and C2 have very little contribution to Fermi level because of their saturated chemical bonds (see also Fig. S3).

**Supplementary Fig. S2. Projected density of states of *Cmcm* structure at phase transition pressure.** Weak hybridization between Ca-$d$ electrons and C-$p$ electrons is observed at Fermi level. Pressure brings about $d$ energy level move to valence bands, which does good to decrease the total energy of system so as to compensate pressure-induced energetic increase, making it be favorable one energetically.

**Supplementary Fig. S3. Electronic local functional (ELF) and structural schematic diagram for *Pmmn* and *Cmcm* phases at phase transitional pressure.** Electronic local functional (ELF)[2] for *Pmmn* structure (a), nanofoam schematic diagram constructed by carbon atoms in *Pmmn* phase, centered with Ca ion chains (b) (unit: Å), Cmcm structure along (100) direction (c), and ELF for *Cmcm* structure (d) and (e). In (b) and (c), inequivalent carbon atoms depicted using different color balls, and calcium atoms shown using green balls. For *Pmmn* structure at 39 GPa, $sp^2$ were connected by $sp^3$ with C-C chemical bonding (length is 1.552 or 1.520 Å). Along y-axis direction, for $sp^2$ carbons, there are two types of C-C bonding with different lengths in zigzag chain (1.467 and 1.426 Å). While for $sp^3$ carbons, the bonding length in zigzag chains along y direction is 1.538 Å. For *Cmcm* structure at 126 GPa, the bonding length could be seen in Table S1.

**Supplementary Fig. S4. Phonon dispersion of *Pmmn* phase at zero pressure obtained using Phonopy code.** Obviously, it is stable dynamically.

**Supplementary Fig. S5. Phonon dispersion of *Cmcm* carbon nanoporous (appeared in phase III) at zero pressure obtained using Phonopy code[3].**

**Supplementary Fig. S6. At zero GPa, band structure of carbon foam (space group *Cmcm*) obtained by removing Ca atoms of phase III (space group, *Pmmn*).** The band gap is about 0.49 eV.

**Supplementary Fig S7. Four types of nanoform structures of intercalation compounds under compression.** $LiC_6$(a), $CaC_6$ (b), and $KC_8$ (c and d).

**Supplementary Fig S8. Total density of states (TDOS) for R-3m at ambient pressure, C222, Pmmn, and Cmcm phases at phase transition pressure.** Obviously, pressure leads to more expanded electronic distribution compared to that of *R-3m*, which leads to lower TDOS in *C222* and *Cmcm*. However, *Pmmn* phase has comparative TDOS with that of *R-3m* phase.

**Supplementary Fig S9. Phonon spectrum of three high pressure phases for $CaC_6$ at cold pressure calculated using Phonopy code.** There are soft modes in peculiar direction for *Pmmn* and *Cmcm* phases.

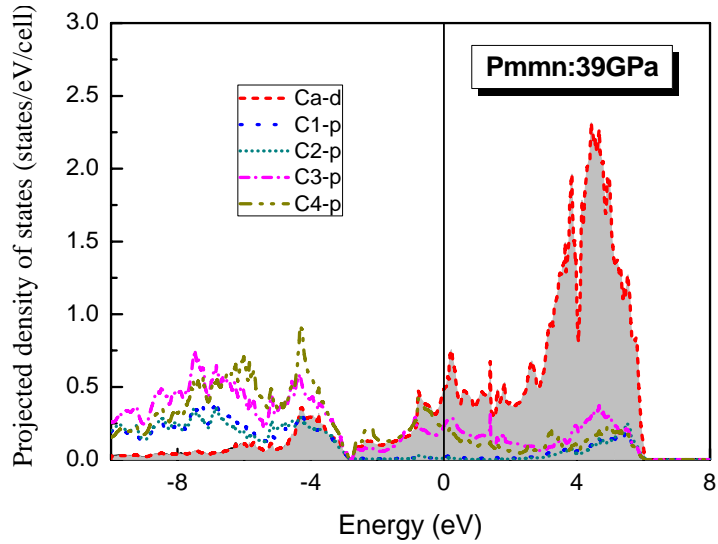

**Supplementary Fig. S1.** Projected density of states of Pmmn structure at phase transition pressure. One can see that there is strong hybridization between Ca-*d* electrons and *p* electrons of C3 and C4 (forming $sp^2$ hybridized zigzag chains along y-axis direction). $sp^3$ hybridized C1 and C2 have very little contribution to Fermi level because of their saturated chemical bonds (see also Fig. S3).

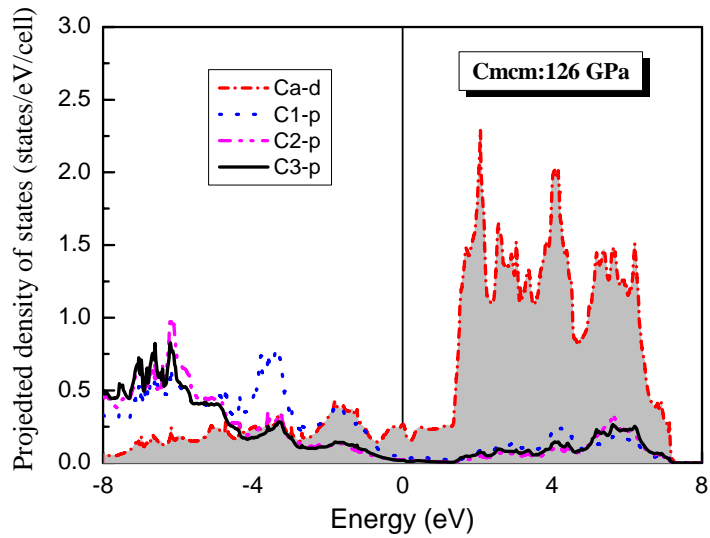

**Supplementary Fig. S2.** Projected density of states of *Cmcm* structure at phase transition pressure. Weak hybridization between Ca-*d* electrons and C-*p* electrons observed at Fermi level. Pressure brings about *d* energy level move to valence bands, which does good to decrease the total energy of system so as to compensate pressure-induced energetic increase, making it be favorable one energetically.

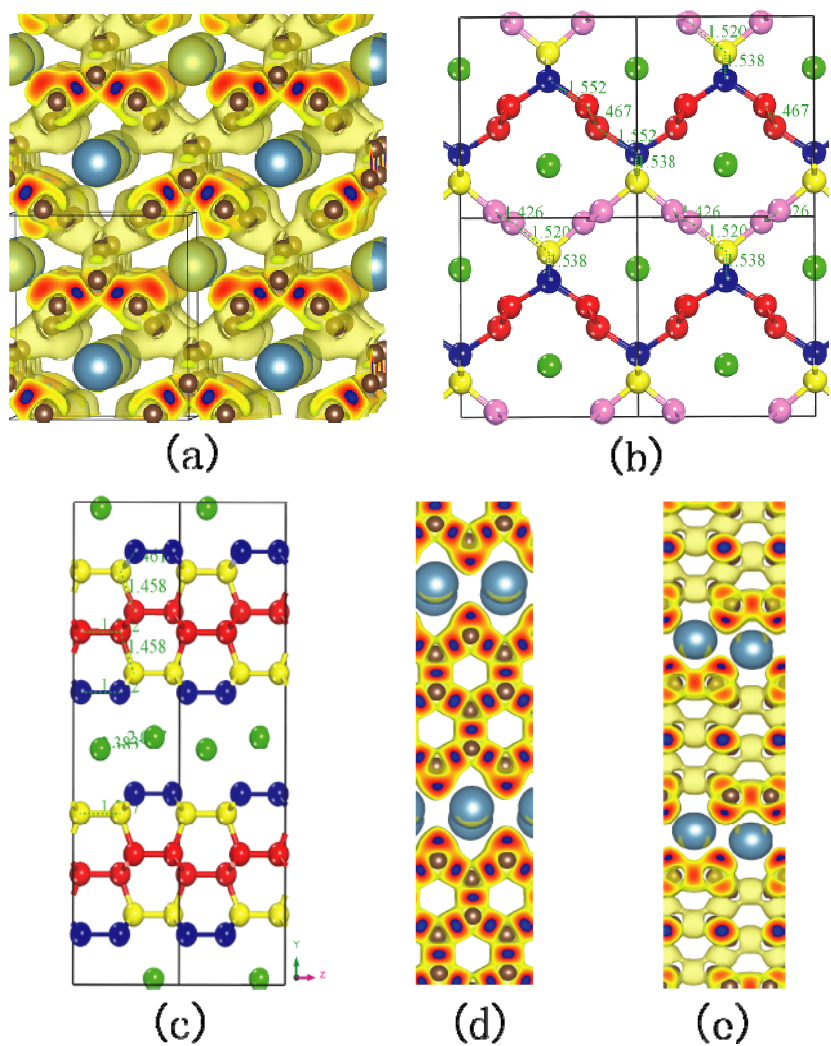

**Supplementary Fig.S3**. Electronic local functional (ELF)$^2$ for *Pmmn* structure (a), nanofoam schematic diagram constructed by carbon atoms in *Pmmn* phase, centered with Ca ion chains (b) (unit: Å), Cmcm structure along (100) direction (c), and ELF for *Cmcm* structure (d) and (e). In (b) and (c), inequivalent carbon atoms depicted using different color balls, and calcium atoms shown using green balls. For *Pmmn* structure at 39 GPa, $sp^2$ were connected by $sp^3$ with C-C chemical bonding (length is 1.552 or 1.520 Å). Along y-axis direction, for $sp^2$ carbons, there are two types of C-C bonding with different lengths in zigzag chain (1.467 and 1.426 Å). While for $sp^3$ carbons, the bonding length in zigzag chains along y direction is 1.538 Å. For *Cmcm* structure at 126 GPa, the bonding length could be seen in Table S1.

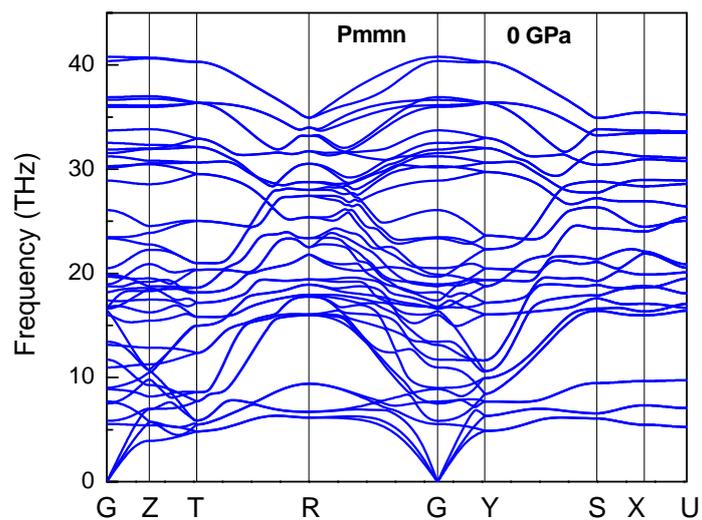

**Supplementary Fig S4.** Phonon dispersion of Pmmn phase at zero pressure obtained using Phonopy code. Obviously, it is stable dynamically.

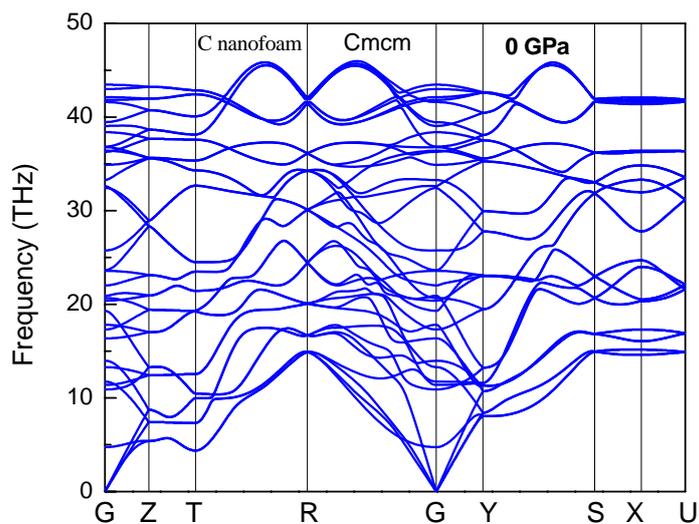

**Supplementary Fig. S5.** Phonon dispersion of Cmcm carbon nanoporous at zero pressure obtained using Phonopy code[3].

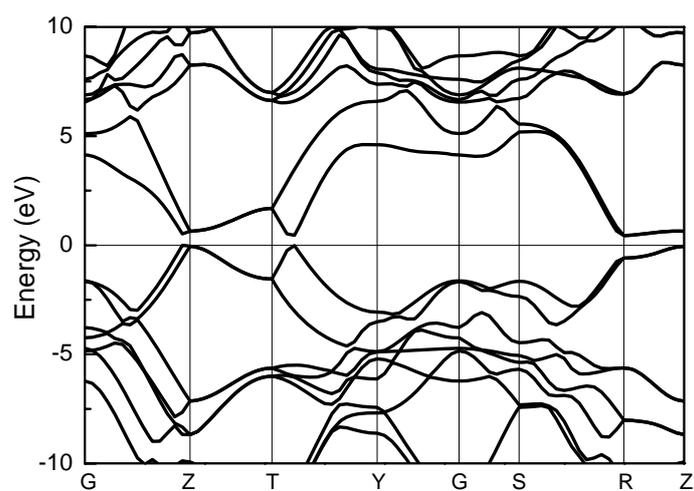

**Supplementary Fig. S6.** At zero GPa, band structure of carbon foam (space group *Cmcm*) obtained by removing Ca atoms of phase III (space group, *Pmmn*). The band gap is about 0.49 eV.

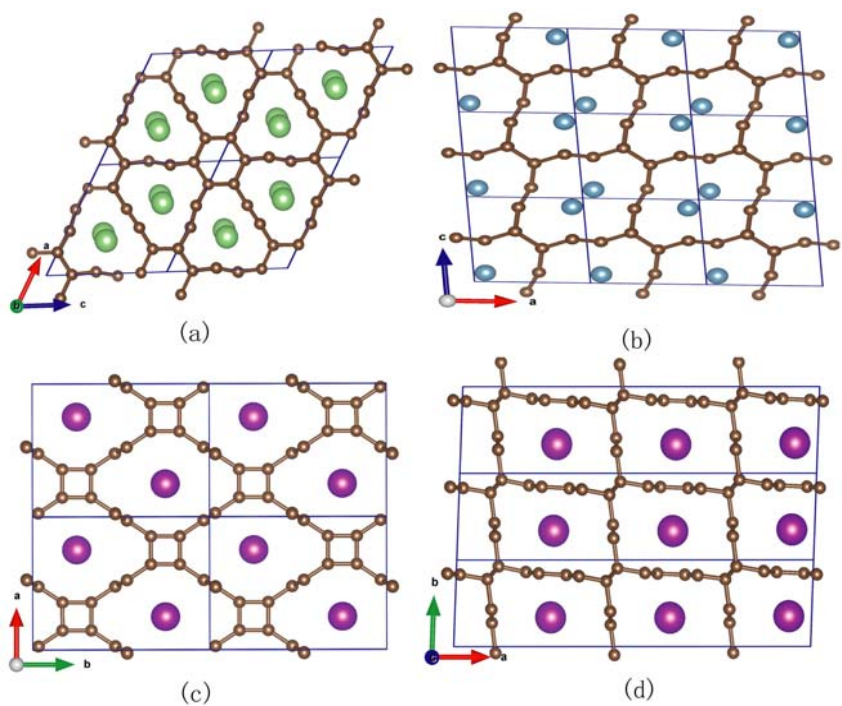

**Supplementary Fig. S7.** Four types of nanoform structures of intercalation compounds under compression. $LiC_6$(a), $CaC_6$ (b), and $KC_8$ (c and d).

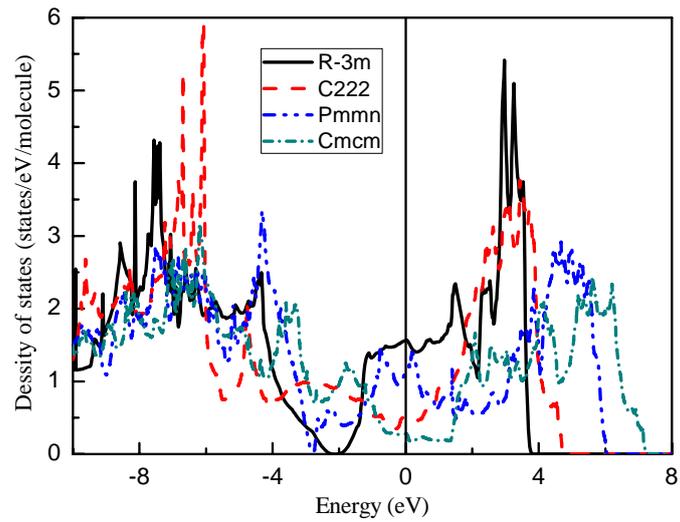

**Supplementary Fig S8.** Total density of states (TDOS) for *R-3m* at ambient pressure, *C222*, *Pmmn*, and *Cmcm* phases at phase transition pressure. Obviously, pressure leads to more expanded electronic distribution compared to that of *R-3m*, which leads to lower TDOS in *C222* and *Cmcm*. However, *Pmmn* phase has comparative TDOS with that of *R-3m* phase.

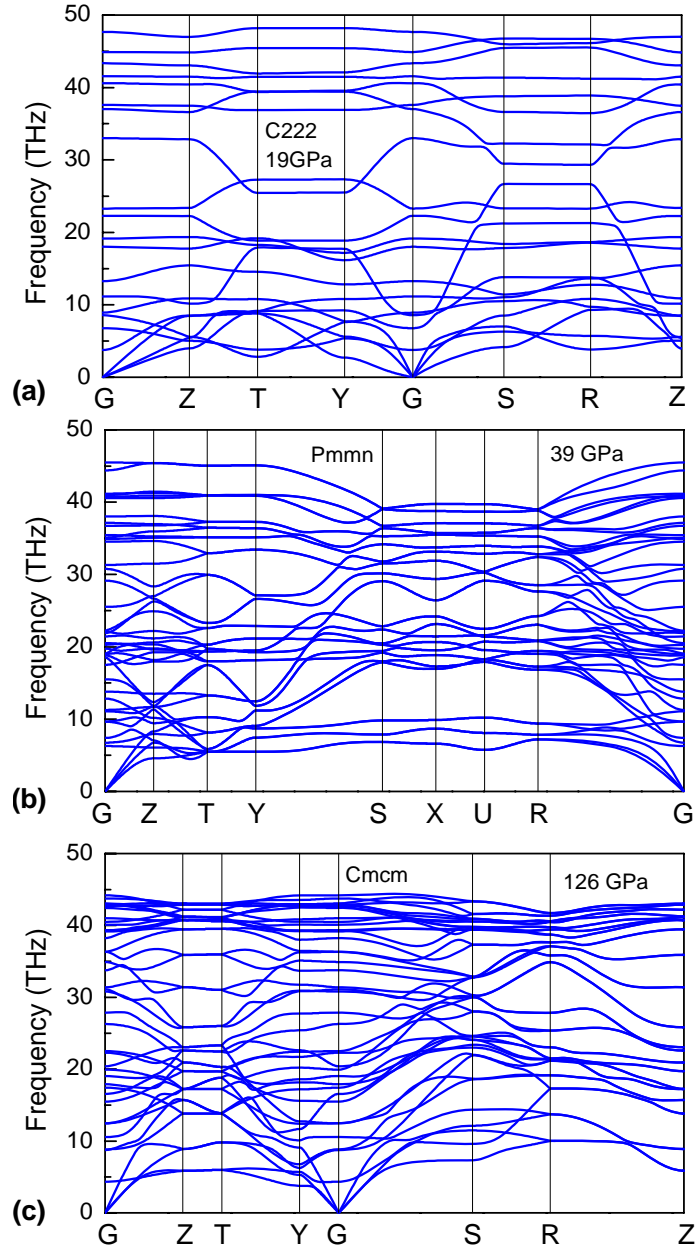

**Supplementary Fig. S9.** Phonon spectrum of three high pressure phases for $CaC_6$ at cold pressure calculated using Phonopy code. There are soft modes in peculiar direction for *Pmmn* and *Cmcm* phases.